\documentclass[aps,pra,a4paper, twocolumn, 10pt]{revtex4-1} 
\usepackage{bbm, amsmath, amssymb, amsthm, bm, textcomp,graphicx,color}
\usepackage[utf8]{inputenc}
\usepackage[T1]{fontenc}
\usepackage[english]{babel}

\theoremstyle{definition}

\begin{document}
\title{Insufficiency of avoided crossings for witnessing large-scale quantum coherence in flux qubits}
\author{Florian Fröwis$^1$, Benjamin Yadin$^{2,3}$, Nicolas Gisin$^1$}
\affiliation{$^{1}$ Group of Applied Physics, University of Geneva, 1211 Geneva, Switzerland\\
  $^2$ QOLS, Blackett Laboratory, Imperial College London, London SW7 2AZ, United Kingdom \\ $^3$ Atomic and Laser Physics, Clarendon Laboratory, University of Oxford, Parks Road, Oxford, OX1 3PU, United Kingdom}
\begin{abstract}
  Do experiments based on superconducting loops segmented with Josephson junctions (e.g., flux qubits) show macroscopic quantum behavior in the sense of Schr\"{o}dinger's cat example? Various arguments based on microscopic and phenomenological models were recently adduced in this debate.
 We approach this problem by adapting --to flux qubits-- the framework of large-scale quantum coherence, which was already successfully applied to spin ensembles and photonic systems. We show that contemporary experiments might show quantum coherence more than one hundred times larger than experiments in the classical regime. However, we argue that the often used demonstration of an avoided crossing in the energy spectrum is not sufficient to conclude about the presence of large-scale quantum coherence. Alternative, rigorous witnesses are proposed.  
\end{abstract}
\date{\today}

\maketitle

\section{Introduction}
\label{sec:introduction}

The experimental demonstration of a massive object in a superposition of two well separated positions is generally considered as a positive test of quantum mechanics on large scales. Typical examples are interference of large molecules or proposed experiments with levitating nanospheres \cite{Arndt_Testing_2014}. These situations are often compared to Schr\"{o}dinger's thought experiment of a cat in a superposition of dead and alive \cite{Schrodinger_gegenwartige_1935}.

\begin{figure}[b!]
\centerline{\includegraphics[width=\columnwidth]{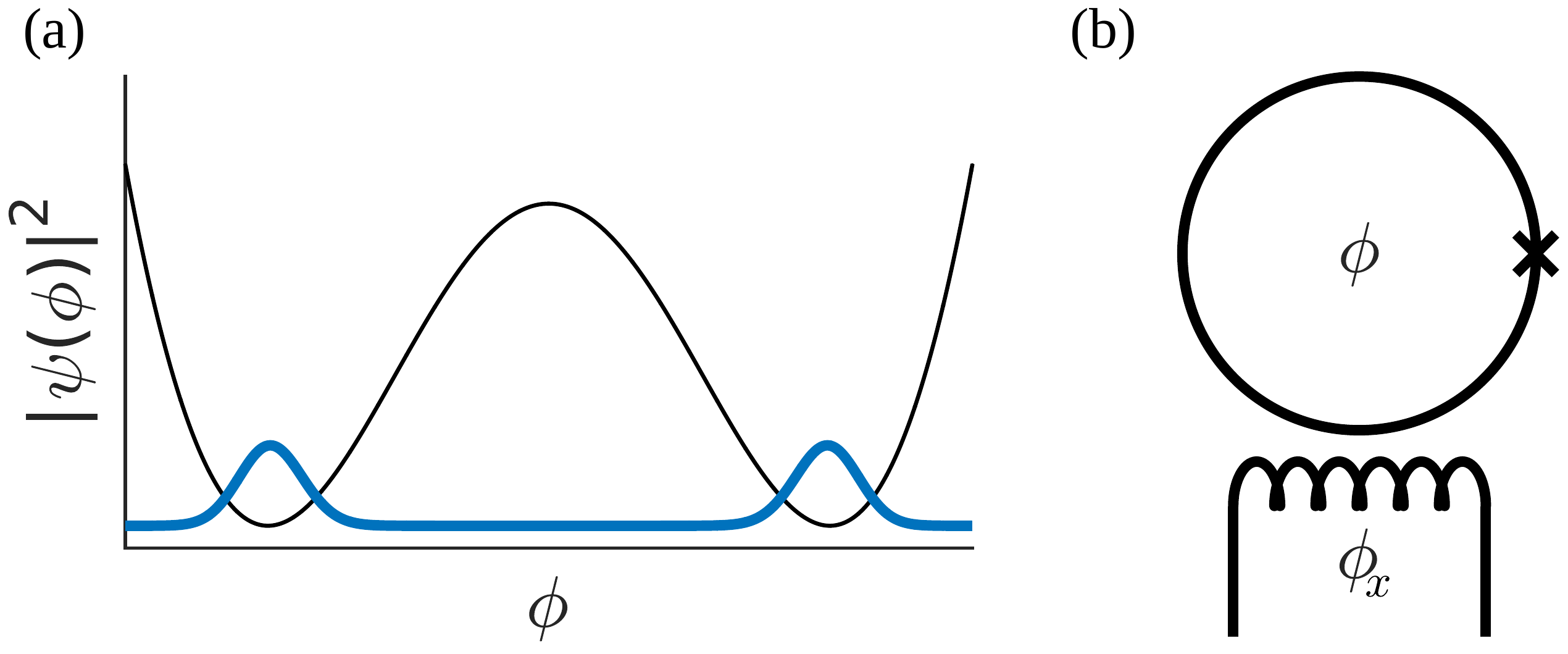}}
\caption[]{\label{fig:schematics} (a) Schematics of a superposition in the flux coordinate $\phi$ in both wells (thick: probability distribution; thin: potential). Experiments witnessing coherence between the two wells are sometimes considered to resemble a Schr\"{o}dinger-cat situation if the inter-well distance and the system size (e.g., number of participating electrons) are large. (b) Most basic schematics of a superconducting ring with a single Josephson junction (cross; see also Sec.~\ref{sec:squid-experiment-}). In a certain parameter regime the magnetic flux $\phi$ through the ring is an appropriate choice for the relevant degree of freedom \cite{Friedman_Quantum_2000,Wal_Quantum_2000,Leggett_Testing_2002,Caldeira_Introduction_2014}. The SQUID is then called flux qubit. The external flux $\phi_x$ controls the effective potential for $\phi$. In practice, the single Josephson junction is replaced by several junctions for an \textit{in situ} control of some experimental parameters \cite{Friedman_Quantum_2000,Wal_Quantum_2000,Hime_Solid-State_2006}.}
\end{figure}

It was argued that superconducting quantum interference devices (SQUIDs), that is, superconducting loops segmented with Josephson junctions, can exhibit a similar characteristic \cite{Leggett_Macroscopic_1980,Leggett_Testing_2002}. In certain parameter regimes, the magnetic flux through the loop can be seen as an appropriate analog to the position of a massive object, where the capacitance of the circuit plays the role of the mass (see Fig.~\ref{fig:schematics}). The nonlinearity of the Josephson junction can lead to an effective double-well potential, in which the minima of the wells correspond to well separated (i.e., ``macroscopically distinct'') flux states. 

There has been a debate about the precise implications of successfully demonstrating a coherent superposition between the two wells  \cite{Leggett_Testing_2002,Korsbakken_Electronic_2009,Korsbakken_size_2010,Leggett_Note_2016,Bjork_size_2004,Marquardt_Measuring_2008,Nimmrichter_Macroscopicity_2013}. On the one hand, it was argued that recent experiments \cite{Friedman_Quantum_2000,Wal_Quantum_2000,Hime_Solid-State_2006} operate in the 100 nA or $\mu$A regime implying the presence of up to $10^{9}$ electrons. This, together with the experimental evidence of ``macroscopic coherence'', should be seen as a genuine ``macroscopic quantum effect'' \cite{Leggett_Testing_2002}. On the other hand, arguments based on microscopic modeling suggest that effectively at most a few thousand electrons make the difference between the two states localized in each well \cite{Korsbakken_Electronic_2009,Korsbakken_size_2010} (see \cite{Leggett_Note_2016} for a critique of this argument). Further contributions also assign an ``effective size'', that is, a number that should, for example, reflect the number of electrons that effectively participate in the observed quantum effect \cite{Bjork_size_2004,Marquardt_Measuring_2008,Nimmrichter_Macroscopicity_2013} (see also table \ref{tab:results}). While the differences in the precise frameworks of \cite{Leggett_Testing_2002,Korsbakken_Electronic_2009,Korsbakken_size_2010,Leggett_Note_2016,Bjork_size_2004,Marquardt_Measuring_2008,Nimmrichter_Macroscopicity_2013} are expected to lead to some deviation in the obtained results, it is astonishing by how much they vary. We find effective sizes ranging from two \cite{Marquardt_Measuring_2008} to 10$^{10}$ \cite{Leggett_Testing_2002}. In addition, none of the theory papers present a conclusive, testable witness for their claim.

In this paper, we argue that one important aspect of the large-scale quantum nature of these experiments is the amount and the spread of quantum coherence in the flux coordinate since it is a direct test of the superposition principle. As already successfully done for spins and photons \cite{Frowis_Measures_2012,Frowis_Linking_2015,Oudot_Two-mode_2015}, we rescale the coherence of the target state by the coherence of a classical reference state (i.e., the state confined in a single well). In this way, we introduce an ``effective size'' that quantifies the scale of the quantum effect. The advantage of this approach is the applicability to experimental data.

After a short review of the motivation and the theory of the framework in Sec.~\ref{sec:conc-large-quant}, the phenomenological model of the flux qubit experiments is presented in Sec.~\ref{sec:squid-experiment-}. Identifying the flux as a relevant observable, we show that indeed the ideally generated target states show large quantum coherence. With the parameters from experiment \cite{Friedman_Quantum_2000}, we find that the quantum coherence of the target states in flux basis is almost two hundred times larger than of a classical reference state (i.e., the ground state of a single potential well). Ideal cat states could even reach numbers 1000 times larger than that of the reference state.

The experimental proof for the generation of the eigenstates was argued to be the avoided crossing in the energy spectrum when tuning the imbalance of the two well minima. In Sec.~\ref{sec:cert-large-quant}, we show that this evidence is not sufficient to conclude the presence of large-scale quantum coherence. We do so by presenting a simple dephasing model in the flux basis that leads to a drastic reduction of the quantum coherence (to the level of the classical reference state) while keeping the feature of the avoided crossing.

The framework of large-scale quantum coherence provides testable
witnesses, which often turn out to be feasible in practice
\cite{Frowis_Lower_2017}. In Sec.~\ref{sec:suff-exper-test}, we
discuss the possibility of lower-bounding large quantum coherence by
witnessing a strong response of the system to flux dephasing. The paper is summarized and discussed in Sec.~\ref{sec:discussion}.

\section{Framework of large quantum coherence}
\label{sec:conc-large-quant}

One of the simplest and most fundamental questions in quantum
mechanics is the validity of the superposition principle on all
scales. There exist several alternative models that, for example,
prevent a persistent superposition of two significantly distinct
positions for massive particles \cite{Bassi_Models_2013}. Given an
observable $X$ with a natural macroscopic limit (such as position or
magnetization), quantum coherence between two far-distant parts of the
spectrum arguably challenges our classical intuition more than quantum
coherence constraint to a small (microscopic) regime \cite{Schrodinger_gegenwartige_1935}.

For pure states, a superposition between far-distant spectral parts of $X$ implies that the wave function has a large spread. The simplest way to measure the spread of a state $| \psi \rangle $ is the variance $\mathrm{var}_{\psi}(X)$. For mixed states, the variance is no longer a faithful measure of coherence since it does not distinguish between coherent superposition and incoherent mixture. The convex roof construction is a well-known technique to overcome the shortcomings of the variance. Given a mixed state $\rho$, one considers the infinite set of all pure state decompositions (PSD) $\rho = \sum_i p_i \left| \psi_i \right\rangle\!\left\langle \psi_i\right| $ and minimizes the average variance 
\begin{equation}
\label{eq:1}
\mathcal{I}(\rho,X) = \min_{\mathrm{PSD}}\sum_i p_i \mathrm{var}_{\psi_i}(X).
\end{equation}
Since the incoherent part is eliminated with the convex roof construction, we call $\mathcal{I}(\rho,X)$ the quantum coherence of $\rho$ in the spectrum $X$. Measuring a certain value of $\mathcal{I}$ experimentally falsifies all collapse models that forbid the superposition principle to be valid on the order of $\sqrt{\mathcal{I}}$. 

Notably, $4 \mathcal{I}$ is the so-called quantum Fisher information \cite{Helstrom_Quantum_1976,Braunstein_Statistical_1994,Yu_Quantum_2013}. This implies that large quantum coherence is necessary to reach high sensitivity in parameter estimation protocols where $X$ is the generator of the parameter shift. The intuitive motivation to choose the convex roof of the variance is made more rigorous in a recent resource theory of ``macroscopic coherence'' \cite{Yadin_General_2016}.

It arguably makes sense to compare $\mathcal{I}(\rho,X)$ to a
reference state $| \psi_{\mathrm{ref}} \rangle $, which behaves
``maximally'' classically among all pure quantum states. Like the
choice for the observable $X$, identifying $| \psi_{\mathrm{ref}}
\rangle $ is physically motivated and hence situation-dependent. Nevertheless it brings some conceptual advantages, as we can avoid unwanted dependencies on scaling factors of $X$. Furthermore, it helps to compare systems with various numbers of modes or particles. One defines 
\begin{equation}
\label{eq:2}
\mathcal{I}_{\mathrm{rel}}(\rho,X) = \frac{\mathcal{I}(\rho,X)  }{\mathcal{I}(\psi_{\mathrm{ref}},X)},
\end{equation}
which tells us how much larger the quantum coherence of $\rho$ is
compared to that of the reference state. 

The observable $X$ and the reference state have to be physically
motivated. Let us consider two examples \cite{Frowis_Measures_2012,Frowis_Linking_2015}. Quadrature operators $X = e^{i\theta} a + e^{-i\theta} a^{\dagger}$ for phase space states (sums of quadrature operators for many modes) and collective spin operators $X = \sum_{i = 1}^N \sigma^{(i)}$ proved to be reasonable choices. The reference states are the coherent state in phase space and the spin-coherent state (i.e., parallel spins in a product state) in spin ensembles, respectively. It was shown that $\mathcal{I}_{\mathrm{rel}}(\rho,X)$ can then be interpreted as an effective size as it is connected to the number of microscopic entities (photons or spins) that effectively contribute to the large-scale quantum effect \cite{Frowis_Macroscopic_2017}. 
Generally, let us consider a classical reference state composed of uncorrelated microscopic constituents (``particles'') and an additive observable with a macroscopic limit. Then the variance is linear in the number of particles $N$, that is, $\mathrm{var}(X) = N \overline{\mathrm{var}}(x_0)$, where $x_0$ is the corresponding one-particle operator and $\overline{\mathrm{var}}$ is the average variance per particle.  Then, the effective size $\mathcal{I}_{\mathrm{rel}}(\rho,X)$ gives the spread of the quantum coherence per particle and per microscopic unit, which justifies the use of the variance rather than, for example, the square root of it.

\section{Short summary of SQUID physics}
\label{sec:squid-experiment-}

\begin{figure}[htbp]
\centerline{\includegraphics[width=\columnwidth]{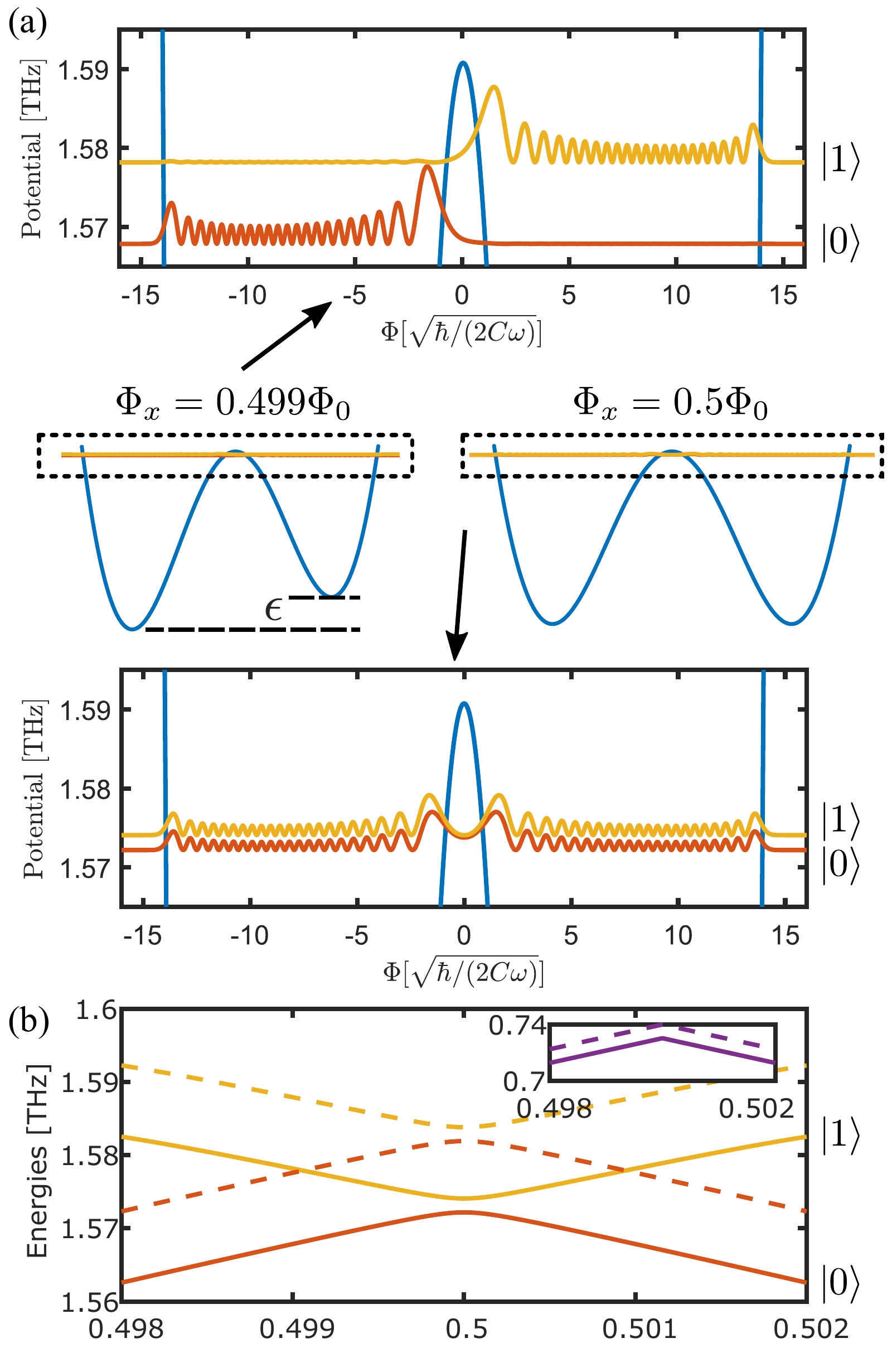}}
\caption[]{\label{fig:friedman} (a) Center: two examples of the double-well potential $U$ (curved, blue line;  Eq.~(\ref{eq:5})) for $\Phi_x = 0.499 \Phi_0$ and $\Phi_x = 0.5 \Phi_0$. The difference $\epsilon$ between the two minima for $\Phi_x = 0.499 \Phi_0$ is artificially increased for a clearer distinction between the two cases. Top: absolute square of the wave functions of the target states $| 0 \rangle $ and $| 1 \rangle $ at $\Phi_x = 0.499 \Phi_0$ in the experiment of Ref.~\cite{Friedman_Quantum_2000}, shifted by the respective eigenenergy. The confinement of each in a single well is clearly visible. Bottom: The same at $\Phi_x = 0.5 \Phi_0$. The wave functions are both spread over both wells. (b) Energies of the target states as a function of $\Phi_x$ (solid lines). The avoid crossing at $\Phi_x = 0.5 \Phi_0$ is clearly visible. The dashed lines are the energies of the target states after dephasing as described in Eq.~(\ref{eq:6}) with $\Gamma = \sqrt{\hbar/(C \omega)}$. The energies are shifted, but the avoided crossing is still present. The inset shows the energy of the ground state before (solid) and after (dashed) dephasing with the same $\Gamma$.\label{fig:EffectDephaseFixCorrLength}\label{fig:WFdetail}}
\end{figure}

In the discussion on whether experiments with superconducting devices show aspects of a Schr\"odinger-cat state, some approaches \cite{Marquardt_Measuring_2008,Korsbakken_Electronic_2009,Korsbakken_size_2010} work with a microscopic model. This leads to discussions about details like the proper choice of the ``microscopic unit'' or of characteristic quantities \cite{Leggett_Note_2016}. In the presented approach based on quantum coherence $\mathcal{I}$, it is more natural to work with a phenomenological model of a single degree of freedom, the total flux $\Phi$ and the charge $Q$ as the conjugate variable (i.e., $[\Phi,Q] = i \hbar$). To discuss a specific example, let us consider the SQUID experiment of Friedman \textit{et al.}~\cite{Friedman_Quantum_2000}. The effective potential depends on the the inductance $L$ and the critical current $I_c$. In addition, it is controlled by an external flux $\Phi_x$ (see Fig.~\ref{fig:schematics} (b)). This results in a quadratic part with inductive energy $E_L = \Phi_0^2/(2L)$ and in a nonlinear part from the junction with Josephson coupling energy $E_J = I_c \Phi_0/(2\pi)$ ($\Phi_0$ is the magnetic flux quantum), that is,
\begin{equation}
\label{eq:5}
U(\phi) = E_L \left( \frac{\Phi-\Phi_x }{\Phi_0} \right)^2 + E_J \cos \left( 2\pi\Phi/\Phi_0 \right).
\end{equation}
The Hamiltonian is completed by the ``kinetic'' energy
$\frac{1}{2C}Q^2$ where $C$ is the capacitance. For later, we
introduce the charging energy $E_C = e^2/(2C)$ and the angular
frequency of the hypothetical $LC$ circuit, $\omega = 1/\sqrt{L
  C}$. The experimental parameters given in
Ref.~\cite{Friedman_Quantum_2000} are $E_C = 188$ MHz, $E_L = 13$ THz
and $E_J = 159$ GHz. In the following we use these numbers when it comes
to numeric calculations, but we generally assume $E_J, E_L
\gg E_C$.

Depending on $\Phi_x$, the potential $U$ can exhibit an effective
double-well structure (see Fig.~\ref{fig:friedman} (a)). At $\Phi_x =
\frac{1}{2} \Phi_0$, there exist two global minima separated by
roughly $0.655\Phi_0$. This implies that coherent tunneling is
possible and the wave functions of the eigenstates have finite
contributions in both wells. This leads to statements like ``the wave
function lives in both wells''. Given the large separation and the
involvement of up to 10$^9$ Cooper pairs, some physicists have called
this a Schr\"odinger-cat situation. The ``classical'' regime is at
$\Phi_x = 0$, as the potential has a (deep) single well with a
well-defined flux state as its ground state. While a small
anharmonicity is always present, the ground state
in this regime behaves like the ground state of
an harmonic oscillator if $E_L, E_J \gg E_C$. For example, the
variance of $\Phi$ and $Q$ are minimal in the sense that the
Heisenberg uncertainty relation is basically tight \footnote{For the
  present energies, we find $\Delta \Phi \Delta Q \approx \hbar/2
  (1+4\times 10^{-7}) \geq \hbar/2$.}. Hence, this state is 
chosen as the reference state as it behaves most classically. 

As argued in Refs.~\cite{Friedman_Quantum_2000,Leggett_Testing_2002}, the experimental proof of the coherent superposition of ``left and right'' is the resolution of the predicted avoided crossing by tuning the imbalance between left and right well (see Fig.~\ref{fig:friedman} (a)). Since the energy gap is proportional to the tunneling probability, the splitting between the lowest two eigenstates is very tiny for the given energies. Hence the initial ground state is driven to highly excited states just below the barrier. There, the minimal energy gap is pronounced enough to be measurable (see Fig.~\ref{fig:friedman} (b)). 

While the discussion in this paper is adapted to the experiment of Ref.~\cite{Friedman_Quantum_2000}, other experimental setups lead to conceptually similar results. For example, the three-junction setup of van der Wal \textit{et al.} \cite{Wal_Quantum_2000} can be modeled with two independent flux coordinates. A suitable coordinate transform leads to an effective double well potential similar to the one discussed before \cite{Orlando_Superconducting_1999}. Hence, we will find qualitative similar results, which are later given without discussing details. We only mention that a further parameter of the experiment is $\alpha$, which is the ratio of the Josephson coupling energy of one junction to the other two (identical) junctions. The parameters from Ref.~\cite{Wal_Quantum_2000} are $E_J/E_C = 38$ and $\alpha = 0.8$.

\section{Large quantum coherence in flux qubits}
\label{sec:appl-superc-devic}

In this section, we apply the framework of large quantum coherence to
the flux qubit. Following on the discussion in
Sec.~\ref{sec:conc-large-quant}, we choose a target state (i.e., the state which is supposed
to exhibit large quantum coherence), an observable $X$ for which we
evaluate the spread of coherence, and a classical reference state to
which we want to compare the target state. From Sec.~\ref{sec:squid-experiment-}, it is evident that eigenstates
(e.g., ground states) at $\Phi_x = 1/2 \Phi_0$ are good candidates for
``cat states'' if we choose $X = \Phi$. The ground
state at $\Phi_x = 0$ serves as the classical reference state. In
Sec.~\ref{sec:effective-size-ideal}, we take the ground state at
$\Phi_x = 1/2 \Phi_0$ as the target state,
while in Sec.~\ref{sec:effect-size-actu} we numerically analyze the
target states of the experiment in
Ref.~\cite{Friedman_Quantum_2000}. Since in both cases we treat the
ideal situation, the quantum coherence $\mathcal{I}$ simplifies to the variance.

\subsection{Effective size of the ideal cat state}
\label{sec:effective-size-ideal}

We first calculate the variance of the classical reference state. For
this, we notice that we are in the regime $E_L/E_C \approx 7.17 \times
10^4 \gg 1$, which implies a rather deep potential in the classical
situation $\Phi_x = 0$. It can be easily shown that the anharmonic
part of the potential does not significantly influence the ground
state (despite a large $E_J$). Hence, to obtain an analytical result,
we approximate $U$ from Eq.~(\ref{eq:5}) by an harmonic oscillator. From the second-order Taylor series of $U$ at $\Phi_x = 0$, we can extract the effective trapping frequency $\omega_{\mathrm{cl}} = \omega \sqrt{1+2\pi^2 E_J/E_L}$ and find 
\begin{equation}
\label{eq:8}
\mathrm{var}_{\mathrm{ref}}(\Phi) \approx \frac{\hbar}{2 C \omega_{\mathrm{cl}}} = \frac{1}{2\pi} \sqrt{\frac{E_C}{E_L + 2\pi^2 E_J}} \Phi_0^2.
\end{equation}

We now turn to the ground state at $\Phi_x = \Phi_0/2$, which is an equally weighted superposition of being in the left and right well. For simplicity, this state is called ``cat state'' in the following. Since the distance $d$ between the minima is much larger than the spread of the wave packet in one well, we approximate the variance of the cat state by $d^2$. In the present parameter regime, $d$ is in the order of $\Phi_0$ with a slight dependency on $E_C$ and $E_J$. We numerically find $\mathrm{var}_{\mathrm{cat}}(\Phi) \approx 0.655^2\Phi_0^2$. Together with Eq.~(\ref{eq:8}), we have 
\begin{equation}
\label{eq:10}
\mathcal{I}_{\mathrm{rel}} (\Phi)\approx 0.86 \pi \sqrt{\frac{E_L + 2\pi^2 E_J}{E_C}} \approx 1315.
\end{equation}
In the limit of a dominant Josephson energy $E_J/E_L \propto I_c
L/\Phi_0 \gg 1$ (while keeping $E_L \gg E_C$), one has $\mathrm{var}_{\mathrm{cat}} (\Phi) \approx \Phi_0^2$ and $\mathcal{I}_{\mathrm{rel}} \approx 2 \sqrt{2}\pi^2 \sqrt{E_J/E_C} \approx 2565$.

The calculation for the experiment of Ref.~\cite{Wal_Quantum_2000} is analogous, but gives a fully analytic expression. We find 
\begin{equation}
\label{eq:11}
\mathcal{I}_{\mathrm{rel}} = 4 \arccos^2 \left( \frac{1}{2\alpha} \right) \sqrt{4\alpha + 2} \sqrt{\frac{E_J}{E_C}} \approx 45.
\end{equation}
For both experiments, we note the scaling of the effective size with $\sqrt{{E_J}/{E_C}}$.

\subsection{Effective size of the target states of Ref.~\cite{Friedman_Quantum_2000}}
\label{sec:effect-size-actu}

We now numerically calculate the effective size of $| 0 \rangle $ and $| 1 \rangle $ (see Fig.~\ref{fig:friedman} (a)) that were targeted in the experiment of Ref.~\cite{Friedman_Quantum_2000}.
For this, we first have to numerically diagonalize the Hamiltonian. As detailed in Ref.~\cite{Everitt_Superconducting_2004}, one analytically calculates the eigenbasis of the Hamiltonian in the case of $E_J = 0$ (since it reduces to the familiar $LC$ circuit) and then expresses the general $H$ in this basis. With the experimental parameters mentioned in Sec.~\ref{sec:squid-experiment-}, we can diagonalize $H$ with high precision and reasonable cut-off dimensions.

Here, we are interested in the regime around $\Phi_x = \frac{1}{2} \Phi_0$. As shown in Fig.~\ref{fig:WFdetail} (a), at $\Phi_x = \frac{1}{2}\Phi_0$, the wave functions are indeed spread over the entire classically allowed range. For the classical reference state, we again use Eq.~(\ref{eq:8}). The eigenstate $| 1 \rangle $ exhibits a spread of $\mathrm{var}_{\left| 1\right\rangle }(\Phi) \approx 6.32 \times 10^{-2} \Phi_0^2$. We find $\mathcal{I}_{\mathrm{rel}} \approx 194$. For the second state $| 0 \rangle $, the spread is slightly smaller.

\begin{table}[tbh]
  \centering
  \caption{\label{tab:results} Comparison of effective size
    $\mathcal{I}_{\mathrm{rel}}$ with results from literature for
    experiments \cite{Friedman_Quantum_2000,Wal_Quantum_2000}. Our
    method is conceptually closest to the work of Bj\"{o}rk and Mana
    \cite{Bjork_size_2004}, whose result has to be squared to match
    ours. Interestingly, we obtain similar numbers as
    \cite{Korsbakken_Electronic_2009} (in the scenario of
    Sec.~\ref{sec:effective-size-ideal}) despite the different
    approach. See Sec.~\ref{sec:discussion} and
    Ref.~\cite{Frowis_Macroscopic_2017} for further discussion.}
  \begin{tabular}{l c c}
    \hline
\hline
Reference& SUNY \cite{Friedman_Quantum_2000} & Delft \cite{Wal_Quantum_2000} \\
    \hline 
    Effective size\footnote{The first number refers to the ideal scenario (Sec.~\ref{sec:effective-size-ideal}), the second to the implemented protocol (Sec.~\ref{sec:effect-size-actu}).} $\mathcal{I}_{\mathrm{rel}}$ & 1315 / 194 & 45 \\
    Leggett\footnote{Leggett uses two measures: disconnectivity and extensive difference.} \cite{Leggett_Testing_2002} & $10^{10} / 10^{10}$ & $10^{6} / 10^{10}$ \\
    Bj\"{o}rk and Mana \cite{Bjork_size_2004} & 33 & - \\
    Marquardt \textit{et al.}~\cite{Marquardt_Measuring_2008} & - & $\lesssim 2$ \\
    Korsbakken \textit{et al.}~\cite{Korsbakken_Electronic_2009} & 3800 - 5750 & 42 \\
\hline
\hline
  \end{tabular}
\end{table}

\section{Certifying large quantum coherence in experiments}
\label{sec:cert-large-quant}

We now turn to the question of how to certify the presence of large quantum coherence from the experimental data. We first show that the demonstration of an avoided crossing in the energy spectrum is not sufficient to witness wide-spread coherence. To this end we introduce a specific dephasing model that largely preserves the energy gap but significantly reduces the quantum coherence. Hence, we discuss alternative schemes at the end of the section.

\subsection{Robustness of avoided crossing, fragility of quantum coherence}
\label{sec:robustn-avoid-cross}

\begin{figure}[t]
\centerline{\includegraphics[width=\columnwidth]{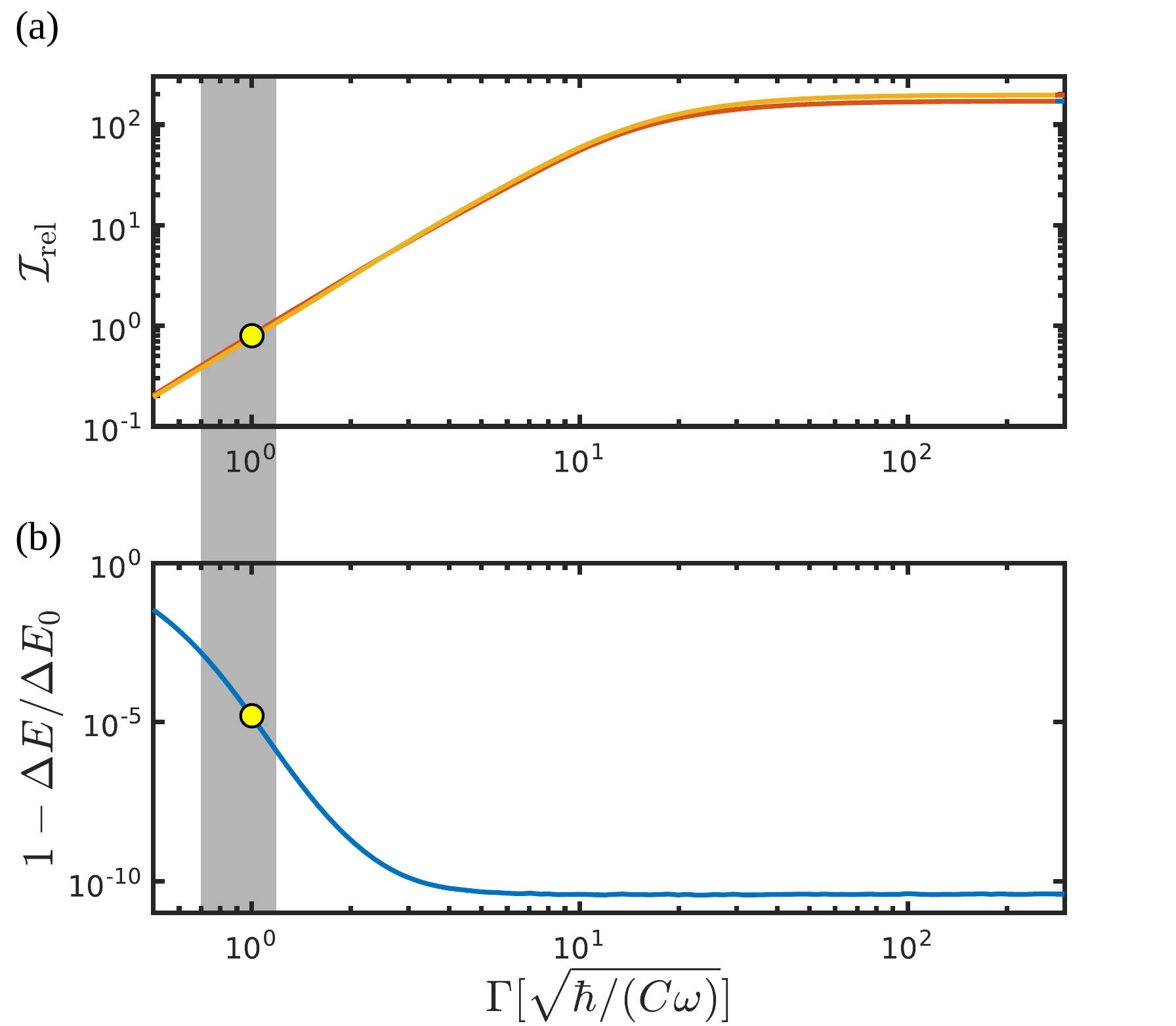}}
\caption[]{\label{fig:EffectDephasingFixPhix} Impact of dephasing as a function of the correlation length $\Gamma$ for the two eigenstates $| 0 \rangle $ and $| 1 \rangle $ (see Sec.~\ref{sec:effect-size-actu} and Fig.~\ref{fig:WFdetail} (a)) at fixed $\Phi_{x} = \frac{1}{2}\Phi_{0}$. (a) The relative quantum coherence $\mathcal{I}_{\mathrm{rel}}$ drops quickly when increasing the dephasing strength (i.e., decreasing $\Gamma$). For $\Gamma \approx 4 \sqrt{\mathrm{var}_{\mathrm{cl}}(\Phi)}$, one has $\mathcal{I}_{\mathrm{rel}} \approx 1$, that is, the spread of the quantum coherence is at the same order as the classical reference state. (b) Relative difference of the energy gap $\Delta E$ under dephasing ($\Delta E_0$ denotes the energy gap without dephasing). The energy gap $\Delta E$ is basically invariant for large $\Gamma$. Only for $\Gamma \lesssim 4 \sqrt{\mathrm{var}_{\mathrm{cl}}(\Phi)}$ the relative difference becomes experimentally relevant. The gray zone highlights the parameter range of $\Gamma$ where $\mathcal{I}_{\mathrm{rel}} \leq 1$ and $1-\Delta E/\Delta E_0 \leq 10^{-3}$. The yellow circles mark $\Gamma = \sqrt{\hbar/(C \omega)}$, which is also used in Fig.~\ref{fig:friedman} (b). The relative difference saturates at around 10$^{-10}$ for large $\Gamma$ because of numerical precision.}
\end{figure}

In this section, we consider a simple dephasing model in the flux basis and discuss its consequences. Suppose that the eigenstates are subject to dephasing
\begin{equation}
\label{eq:6}
\rho(\Phi,\Phi^{\prime}) \rightarrow e^{- (\Phi-\Phi^{\prime})^2/ (2\Gamma^2)}\rho(\Phi,\Phi^{\prime}),
\end{equation}
meaning that coherence-elements $\rho(\Phi,\Phi^{\prime})$ are damped by a Gaussian function on a scale $\Gamma$ (called correlation length in the following). From the discussion in Sec.~\ref{sec:conc-large-quant} it is not surprising that this map reduces the quantum coherence $\mathcal{I}$, which is calculated using the spectral decomposition $\rho = \sum_i \lambda_i \left| \psi_i \right\rangle\!\left\langle \psi_i\right| $ and 
\begin{equation}
\label{eq:7}
\mathcal{I}(\rho,\Phi) = \sum_{i < j} \frac{(\lambda_i - \lambda_j)^{2}}{\lambda_i + \lambda_j} \left| \left\langle \psi_i \right| \Phi \left| \psi_j \right\rangle  \right|^2
\end{equation}
(see, e.g., Ref.~\cite{Braunstein_Statistical_1994}). The result for $\Phi_x = \frac{1}{2}\Phi_{0}$ and variable $\Gamma$ is presented in Fig.~\ref{fig:EffectDephasingFixPhix}, which shows the fragility of the quantum coherence under flux dephasing. In contrast, the energy gap $\Delta E$ (i.e., the difference in the energy expectation value between the dephased eigenstates at $\Phi_x = \frac{1}{2}\Phi_0$) is rather stable (see Fig.~\ref{fig:EffectDephasingFixPhix} (b)). Only for a correlation length $\Gamma \lesssim \sqrt{\hbar/(C \omega)}$ (i.e., close to the width of the classical reference state) $\Delta E$ deviates significantly from the energy gap between the original eigenstates, denoted by $\Delta E_0$. Note that the energies of the states are increased by the map (\ref{eq:6}). Since this affects all states equally strong (including the ground state), the differences in energies remain basically unchanged.

From these findings, we conclude that the mere observation of an avoided crossing is not sufficient to prove large quantum coherence in the system. We emphasize that this conclusion cannot be found by reducing the flux coordinate to a two-dimensional space (left well/ right well), which is often done in qualitative discussions.

\subsection{Sufficient experimental test}
\label{sec:suff-exper-test}

We now discuss the possibility for a sufficient experimental test to certify a minimal quantum coherence $\mathcal{I}(\rho,\Phi)$. As shown in Ref.~\cite{Frowis_Detecting_2016}, we can employ the following dynamical protocol. For this, we have to be able to implement $V_t = \exp(-i \Phi t)$. We first choose a suitable measurement with measurement operators $\{\Pi_k\}$ for the outcomes $\{x_k\}$. Then, we measure the probability distributions $p_k = \mathrm{Tr}\rho \Pi_k$ and $q_k = \mathrm{Tr}V_t\rho V_t^{\dagger} \Pi_k$. With this, we calculate the so-called Bhattacharyya coefficient $B = \sum_k \sqrt{p_k q_k}$. One finds \cite{Frowis_Detecting_2016}
\begin{equation}
\label{eq:4}
\mathcal{I}(\rho,\Phi) \geq \frac{1}{t^2} \arccos^2 B.
\end{equation}
In essence, inequality Eq.~(\ref{eq:4}) witnesses the response of the system to a dynamical process generated by $\Phi$. High susceptibility implies the presence of large-scale quantum coherence. Using this bound, one finds certified quantum coherence which is (in the cold atom experiment of Ref.~\cite{Hosten_Measurement_2016}, for example) up to $71$ larger than that of classical reference states \cite{Frowis_Lower_2017}.

As we show now, we can relax the requirements of this protocol. For instance, suppose we are not able to fully control $t$ (i.e., the duration of the unitary), but we only assume that $t$ follows a certain distribution $\mu(t)$ with $\int dt \mu(t) = 1$. Then, the quantum state after the non-unitary dynamics is described by 
\begin{equation}
\label{eq:9}
\mathcal{E}(\rho) = \int_{-\infty}^{\infty} dt \mu(t) V_t \rho V_t^{\dagger}.
\end{equation}
For example,
\begin{equation}
\mu(t) = \frac{1}{\sqrt{2\pi} \Gamma} \exp(- \Gamma^2 t^2/2)\label{eq:15}
\end{equation}
leads to an effective dephasing in the $\Phi$ basis exactly as modeled in Eq.~(\ref{eq:6}). We redefine $q_k = \mathrm{Tr} \mathcal{E}(\rho) \Pi_k$ for the calculation of $B$. Following Ref.~\cite{Frowis_Detecting_2016} and using the quantum fidelity $F$, one can show that 
\begin{equation}
  \begin{split}
    B &\geq F(\rho, \mathcal{E}(\rho)) \geq \int_{-\infty}^{\infty} dt
    \mu(t) F(\rho, V_t \rho V_t^{\dagger})\label{eq:14}
    \\
    & \geq \int_{- \pi/(2\sqrt{\mathcal{I}})}^{
      \pi/(2\sqrt{\mathcal{I}})} dt \mu(t) \cos \left(
      \sqrt{\mathcal{I}} t \right),
  \end{split}
\end{equation}
where for the last step we used $F(\rho, V_t \rho V_t^{\dagger}) \geq
\cos \left( \sqrt{\mathcal{I}} t \right)$ in the interval $t \in [-
\pi/(2\sqrt{\mathcal{I}}), \pi/(2\sqrt{\mathcal{I}})]$
\cite{Uhlmann_gauge_1991,Frowis_Kind_2012}. Finally, one has to invert
inequality (\ref{eq:14}) to obtain a lower bound on the quantum
coherence $\mathcal{I}$. This is generally not analytically possible,
but one can find either numerical solutions or approximations. The
numerical solution is an optimization problem where one finds the
maximal $\mathcal{I}$ such that the inequality in Eq.~\eqref{eq:14} is fulfilled given the data, $B$, and the
model of the dynamics, $\mu(t)$. As an example for an analytical
approximation, we consider Eq.~(\ref{eq:15}). In the weak dephasing
limit (i.e., $\Gamma^2 \gg \mathcal{I}$), we can safely move the
integration limits in the second line of Eq.~\eqref{eq:14} from $\pm \pi/(2 \sqrt{\mathcal{I}})$ to
$\pm \infty$, which allows us to find $B \gtrsim \exp (-\Gamma^2
\mathcal{I}/2)$, from which 
\begin{equation}
\label{eq:12}
\mathcal{I}(\rho,\Phi) \gtrsim 2\Gamma^2 \log 1/B
\end{equation}
follows.

Recent experiments with flux qubits (e.g., by Knee \textit{et al.}
\cite{Knee_strict_2016a}) are conceptually and technically very close
to the presented framework. A standard way of implementing
measurements is based on the Josephson Bifurcation Amplifier
\cite{Siddiqi_RF-Driven_2004,Lupascu_Quantum_2007,Nakano_Quantum_2009},
which can be modeled by a dephasing process similar to
Eq.~(\ref{eq:9}). This opens the opportunity for witnessing large
quantum coherence with flux qubits. Then, the protocol in
Ref.~\cite{Knee_strict_2016a}, which was employed to falsify so-called
macrorealistic theories, is of the same spirit as the procedure
presented here, namely to witness the susceptibility of a system with
respect to a specific influence. The difference is that here we do
need to control the experiment to a certain degree. In particular, the
dynamics of Eq.~(\ref{eq:9}) should be well characterized (e.g.,
$\Gamma$ in Eq.~(\ref{eq:15}) should be sufficiently known), while
assumptions on the initial state $\rho$ and the measurement
$\{\Pi_k\}$ are not necessary.

\section{Discussion and summary}
\label{sec:discussion}

In this paper, we proposed to consider the extent of quantum coherence in the flux coordinate as one relevant aspect of the ``macroscopic quantum nature'' of flux qubits. While the framework of large quantum coherence arguably has an operational meaning in the resource theory of asymmetry \cite{Marvian_How_2016,Yadin_General_2016} and metrology \cite{Helstrom_Quantum_1976,Braunstein_Statistical_1994}, its interpretation as an ``effective size'' (i.e., how many Cooper pairs or electrons effectively contribute to the quantum phenomenon) is less clear.

From a mathematical point of view, the framework of large quantum coherence is directly applicable to the SQUID experiments. The choice of observable (here, the flux $\Phi$) and the reference state (ground state of the ``classical'' regime $\Phi_x = 0$) are physically motivated. The scaling of the effective size in the ideal case (Sec. \ref{sec:effective-size-ideal}) with $\sqrt{E_J/E_C}$ is interesting as the ratio $E_J/E_C$ is sometimes argued to be the relevant scale of the experiment \cite{Marquardt_Measuring_2008}.

In this context, it is interesting to compare our results with the analysis by Korsbakken \textit{et al.} \cite{Korsbakken_Electronic_2009,Korsbakken_size_2010} \footnote{Further discussion regarding the theoretical estimation of the ``effective size'' of the SQUID done in other papers is provided in Ref.~\cite{Frowis_Macroscopic_2017}.}. The authors work with a microscopic model to investigate the effective number of participating Cooper pairs in the cat state. They find numbers around 3800 - 5750 for Ref.~\cite{Friedman_Quantum_2000} and 42 for Ref.~\cite{Wal_Quantum_2000}. Since their results depend on experimentally ``difficult'' quantities like the Fermi velocity, this work was criticized by Leggett in Ref.~\cite{Leggett_Note_2016} where he constructs an ``obvious'' example of a Schr\"odinger-cat state that still gives a very low effect size with Korsbakken \textit{et al.}'s method. Nevertheless, we note that our results (which only depend on the phenomenological model) and the results of Korsbakken \textit{et al.} are in the same order. It would be interesting to see whether this is pure coincidence or whether there is some underlying connection.

We have shown that the experimental observation of avoided crossing in
the energy spectrum is not a conclusive witness for the presence of
large-scale quantum coherence. In contrast, the framework of large
quantum coherence provides powerful lower bounds that are testable in
practice. Here, we extended a recently developed protocol
\cite{Frowis_Detecting_2016} to nonunitary dynamics, which is
conceptually and technically close to existing experiments
\cite{Knee_strict_2016a}. Note that Eq.~\eqref{eq:14} is a general way
to lower bound the spread of quantum coherence for an observable $X$. For its successful
application, it is necessary to faithfully implement the unitary $V_t =
\exp(-i t X)$ in the experiment and to observe the impact by measuring
the system before and after its application (on different runs). The control of $t$ can be
relaxed to having knowledge about its distribution over many runs,
$\mu(t)$. In contrast, no knowledge about the measurement is necessary
to guarantee the correctness of Eq.~\eqref{eq:14}.

The presented work gives a first insight that wide-spread
quantum coherence might be present in SQUID experiments. We see our results as a starting
point for further investigations. As mentioned
before, the microscopical interpretation of
$\mathcal{I}_{\mathrm{rel}}$ is open. Furthermore, one should apply
the framework of large quantum coherence to other experiments and
parameter regimes, such as $E_J > E_C > E_L$ (i.e., fluxonium) and
$E_J \approx E_L \gg E_C$ (c-shunt flux qubit). A well-justified
choice of the relevant observable as well as the target and reference
state will lead to further insight to meso- and macroscopic quantum phenomena.

\textit{Acknowledgments.---} We thank Daniel Estève and his research team, Amir Caldeira, Bill Munro, George Knee, Pavel Sekatski and Wolfgang Dür for inspiring discussions. This work was supported by the National Swiss Science Foundation (SNF), grant No. 172590, the European Research Council (ERC MEC), the EPSRC and Wolfson College, University of Oxford.

\bibliographystyle{apsrev4-1}
\bibliography{SQUID}
\end{document}